\documentclass[fleqn,12pt,twoside]{article}
\usepackage{espcrc1}
\usepackage{epsfig}

\usepackage{graphicx}
\usepackage[figuresright]{rotating}

\newcommand{\AmS}{{\protect\the\textfont2
  A\kern-.1667em\lower.5ex\hbox{M}\kern-.125emS}}

\title{Hyperspherical harmonic study of identical-flavor four-quark systems}

\author{J. Vijande \address{Dpto. de F\'\i sica Te\'orica 
        and IFIC, Universidad de Valencia - CSIC, Spain},
        N. Barnea\address{The Racah Institute of Physics, 
	The Hebrew University,
	Jerusalem, Israel},
	and
        A. Valcarce\address{Grupo de F\'\i sica Nuclear and IUFFyM,
        Universidad de Salamanca, Spain}}
       
\begin{document}

% typeset front matter
\maketitle

\begin{abstract}
We present an exact method based on a hyperspherical harmonic expansion 
to study systems made of quarks and antiquarks of the same flavor. 
Our formalism reproduces and improves the results obtained 
with variational approaches.
This analysis shows that identical-flavor four-quark systems with non-exotic
$2^{++}$ quantum numbers may be bound independently of the quark mass.
$0^{+-}$ and $1^{+-}$ states become attractive only for larger quarks masses.
\end{abstract}

\section{Introduction}

The understanding of few-body systems relies in our capability to design
methods for finding an exact or approximate solution of the $N-$body problem.
The solution of any few-particle system may be found 
by means of an expansion of the trial wave function in terms 
of hyperspherical harmonic (HH)
functions. The idea is to generalize the simplicity of the spherical harmonic
expansion for the angular functions of a single particle motion to a system of
particles by introducing a global length $\rho$, called the hyperradius, and a
set of angles, $\Omega$. For the HH expansion method to be practical, the
evaluation of the potential energy matrix elements must be feasible. The main
difficulty of this method is to construct HH functions of proper symmetry for a
system of identical particles. This may be
overcome by means of the HH formalism based on the symmetrization of the
$N-$body wave function with respect to the symmetric group using the Barnea and 
Novoselsky algorithm\cite{Nir9798}. 

The recent discoveries of several meson--like resonances 
whose properties do not fit into the predictions of the
naive quark model, has reopened the interest on the possible role
played by non-$q\bar q$ configurations in the meson 
spectra. Among them, the possible existence of 
bound four-quark states (two quarks and two antiquarks) 
has been suggested in the low-energy hadron 
spectroscopy. Four-quark bound states were already suggested 
theoretically thirty years ago, both in the light-quark
sector by Jaffe\cite{Jaf77} and in the heavy-quark sector by Iwasaki\cite{Iwa76}. 

In this work we present a general study of four-quark systems of identical flavor
in an exact way. For this purpose we have generalized the HH method, 
widely used in traditional nuclear physics for the study of
few-body nuclei, to describe systems made of quarks and antiquarks. 
This generalization presents two main difficulties,
first the simultaneous treatment of particles and antiparticles, and second
the additional color degree of freedom. 

\section{General formulation of the problem}

The system of two quarks and two antiquarks with the same flavor can be regarded
as a system of four identical particles. Each particle carries
a $SU(2)$ spin label and a $SU(3)$ color label. Both quarks and antiquarks are
spin $1/2$ particles, but whereas a quark color state
belongs to the $SU(3)$ fundamental representation 
$[3]$, an antiquark color state is a member of the fundamental 
representation $[\bar 3]$.   
The four-body wave function is a sum of outer products
of color, spin and configuration terms
\begin{equation}
    | \phi \rangle = |{\rm Color}\rangle |{\rm Spin}\rangle | R \rangle
\end{equation}
coupled to yield an antisymmetric wave function with
a set of quantum numbers that reflects the symmetries of the system.
These are the total angular momentum quantum number $J$, its projection $J^z$, 
and the $SU(3)$ color state $G$ (labeled as $G$ to avoid 
confusion with the charge conjugation quantum number),
which by assumption must belong for physical states
to the $SU(3)$ color singlet representation. 
Since QCD preserves parity, parity is also a good 
quantum number. Another relevant quantum number to the system
under consideration it is the $C-$parity, $C$, 
i.e., the symmetry under interchange of quarks and antiquarks.

To obtain a solution of the four-body Schr\"odinger equation we 
eliminate the center of mass and use the relative, Jacobi,
coordinates $\vec{\eta}_1,\vec{\eta}_2,\ldots,\vec{\eta}_{A-1}$. Then
we expand the spatial part of the 
wave-function using the HH basis. In this formalism the Jacobi coordinates 
are replaced by one radial coordinate,  
the hyperradius $\rho$, and a set of ($3A-4$) angular coordinates 
$\Omega_A$. The HH basis functions are eigenfunctions of the hyperspherical
part of the Laplace operator.
An antisymmetric $A$--body basis functions 
with total angular momentum $J_A,J^z_A$, color $G_A $
and $C$-parity $C$, are given by,
\begin{eqnarray} \label{HH_A}
| n {K}_A J_A J^z_A G_A C \Gamma_A \alpha_A \beta_A \rangle  & = &
\cr & & \hspace{-25mm}
      \sum_{Y_{A-1}}
      \frac{\Lambda_{\Gamma_{A},Y_{A-1}}}{\sqrt{| \Gamma_{A}|}} \,
      \left[ | K_A L_A M_A \Gamma_A Y_{A-1} \alpha_A \rangle
             | S_A S^z_A G_A C 
               \, \widetilde{\Gamma}_{A},\widetilde{Y}_{A-1}
                  \, \beta_A \rangle
          \right]^{J_A J^z_A} | n \rangle \,,
\end{eqnarray}
where $\langle \rho | n \rangle \equiv R_n(\rho)$
are the hyperradial basis functions, taken to be Laguerre functions.
\begin{equation}
\langle \Omega_A | K_A L_A M_A \Gamma_A Y_{A-1} \alpha_A \rangle
      \equiv 
{\cal Y}^{[A]}_{K_A L_A M_A \Gamma_A Y_{A-1} \alpha_A}(\Omega_A)   
\end{equation}
are HH functions with hyperspherical angular momentum $K=K_A$, 
and orbital angular momentum quantum numbers ($L_A, M_A$) that belong
to well-defined irreducible representations (irreps)
$\Gamma_{1} \in \Gamma_2 \ldots \in \Gamma_A $ of the permutation 
group--subgroup chain  
${\cal S}_1 \subset {\cal S}_2 \ldots \subset {\cal S}_A $,
denoted by the Yamanouchi symbol 
$[ \Gamma_A, Y_{A-1} ] \equiv [ \Gamma_A,\Gamma_{A-1},\ldots,\Gamma_1 ]$.
The dimension of the irrep $\Gamma_{m}$ is denoted by $| \Gamma_{m}|$
and $\Lambda_{\Gamma_{A},Y_{A-1}}$ is a phase factor.
Similarly, the functions 
\begin{equation}
   \langle s^z_1..s^z_A, g_1..g_A 
   |  S_A S^z_A G_A C \, \widetilde{\Gamma}_{A},\widetilde{Y}_{A-1} 
                                                      \beta_A \rangle
   \equiv 
   \chi^{[A]}_{S_A S^z_A G_A \, \widetilde{\Gamma}_{A},\widetilde{Y}_{A-1}
                  \, \beta_A}(s^z_1..s^z_A, g_1..g_A)
\end{equation}
are the symmetrized color--spin basis functions, given in terms of the
spin projections ($s^z_i$) and color states ($g_i$) of the particles.
The quantum numbers $\alpha_A$, and $\beta_A$ are used to remove the degeneracy 
of the HH and color--spin states, respectively. For the construction
of the symmetrized HH basis we use the algorithm of Barnea and 
Novoselsky \cite{Nir9798}, which utilizes the group of kinematic rotations.
For the color--spin subspace, we use a method to transform the standard basis
into a symmetrized color--spin basis with well defined color and $C-$parity.
The technical steps of such construction are detailed in Ref. \cite{Bar06}.
The calculation of the Hamiltonian matrix-elements between the 
antisymmetric basis functions, Eq. (\ref{HH_A}), is almost the same
as in nuclear physics, replacing isospin by color \cite{BLO99}.

\begin{table}[htb]
\caption{$cc\bar c\bar c$ masses for the maximum value of $K$ used, $E(K_{max})$, and using 
the extrapolation, $E(K=\infty)$, compared to the corresponding threshold,
$M_1M_2$ and $T(M_1,M_2)$. The subindex stands for final state relative angular momentum. 
The value of $\Delta$ for each state is also given. Energies are in MeV.}
\label{t2}
\newcommand{\m}{\hphantom{$-$}}
\newcommand{\cc}[1]{\multicolumn{1}{c}{#1}}
\renewcommand{\tabcolsep}{1.5pc} % enlarge column spacing
\renewcommand{\arraystretch}{1.2} % enlarge line spacing
\begin{tabular}{@{}|c|c|c|c|} \hline
\hline
$J^{PC}$  & $E(K=\infty)$ [$E(K_{max})$]& $M_1M_2$ [$T(M_1,M_2)$] & $\Delta$\\
\hline
$ 0^{++}$ & 6038 [6115] & $\eta_c\,\eta_c\vert_S$  [5980] & +58  \\ 
$ 0^{+-}$ & 6515 [6606] & $\eta_c\,h_c\vert_P$	   [6497] & +18 \\
$ 1^{++}$ & 6530 [6609] & $\eta_c\,\chi_{c0}\vert_P$ [6433] & +97 \\
$ 1^{+-}$ & 6101 [6176] & $J/\psi\,\eta_c\vert_S$  [6087] &  +14 \\
$ 2^{++}$ & 6172 [6216] & $J/\psi\,J/\psi\vert_S$  [6194] &  $-$22 \\
$ 2^{+-}$ & 6586 [6648] & $\eta_c\,h_c\vert_P$     [6497] & +89 \\
$ 0^{-+}$ & 6993 [7051] & $J/\psi\,J/\psi\vert_P$  [6194] & +779 \\
$ 0^{--}$ & 7276 [7362] & $J/\psi\,\eta_c\vert_S$  [6087] & +1189 \\
$ 1^{-+}$ & 7275 [7363] & $J/\psi\,J/\psi\vert_P$  [6194] & +1081 \\
$ 1^{--}$ & 6998 [7052] & $J/\psi\,\eta_c\vert_S$  [6087] & +911 \\
$ 2^{-+}$ & 7002 [7055] & $J/\psi\,J/\psi\vert_P$  [6194] & +808 \\
$ 2^{--}$ & 7278 [7357] & $J/\psi\,\eta_c\vert_S$  [6087] & +1191 \\
\hline
\end{tabular}
\end{table}

\section{Results}

We have applied our method to study $cc\bar c\bar c$ states. The calculation
has been done up to the maximum value of $K$ within our computational
capabilities ($K_{max}$). 
To analyze the stability of these systems against dissociation through strong decay,
parity ($P$), $C-$parity
($C$), and total angular momentum ($J$) must be conserved. 

We show in Table \ref{t2} all possible $J^{PC}$ quantum numbers with $L=0$. 
We also indicate the lowest two-meson threshold for each set of
quantum numbers. Let us note that the convergence of the expansion in terms of hyperspherical
harmonics is slow, and the effective potential techniques\cite{BLO99} do not improve it.
To obtain a more adequate value for the energy we have extrapolated it according 
to $ E(K)=E(K=\infty)+a/K^b$, where $E(K=\infty)$, $a$ and $b$ are fitted parameters. 
The values obtained for $E(K=\infty)$ are stable within $\pm$10 MeV for each quantum number.
Four-quark states will be stable under strong 
interaction if their total energy lies below all possible, and allowed, two-meson 
thresholds. It is useful to define $\Delta=E(K=\infty)-T(M_1,M_2)$
in such a way that if $\Delta>0$ the four-quark system will fall
apart into two mesons, while $\Delta<0$ will indicate that such strong decay is
forbidden and therefore the decay, if
allowed, must be weak or electromagnetic, being its width much narrower.

A first glance to these results indicates that 
only three sets of quantum numbers have some probability of being observed.
These are the $J^{PC}=0^{+-}$, $1^{+-}$, and $J^{PC}=2^{++}$, 
which are very close to the corresponding threshold.
It is interesting to observe that the quantum numbers $0^{+-}$
correspond to an exotic state, those whose quantum numbers cannot be obtained from a
$q\bar q$ configuration. 

To analyze whether the existence of bound states with exotic quantum numbers could be a characteristic 
feature of the heavy quark sector or it is also present in the light sector we have calculated the value of 
$\Delta$ for different quark masses.
We have obtained that only one of the non-exotic states, the $2^{++}$, becomes more bound when the quark mass is
decreased, $\Delta\approx-80$ MeV for the light quark mass. The $1^{+-}$ and $0^{++}$
states, that were slightly above threshold in the charm sector, increase
their attraction when the quark mass is increased and only for masses close to the bottom quark mass may be bound. 
With respect to the
exotic quantum numbers, the negative parity $0^{--}$ and
$1^{-+}$ are not bound for any value of the quark mass. 
Only the $0^{+-}$ four-quark state becomes more deeply bound when the constituent quark mass increases,
and therefore only one possible narrow state with exotic quantum numbers may appear in the heavy-quark sector. This 
state presents an open $P-$wave threshold only for quark masses below 3 GeV.

Let us note that since the heavy quarks are isoscalar states, the flavor wave function of the four heavy-quark
states will be completely symmetric with total isospin equal to zero. Therefore, one should 
compare the results obtained in the light-quark case with a completely symmetric flavor 
wave function, i.e., the isotensor states.
There are experimental evidences for three states with exotic quantum numbers 
in the light-quark sector. Two of them are isovectors with quantum numbers $J^{PC}=1^{-+}$ named
$\pi_1(1400)$ and $\pi_1(1600)$, 
and one isotensor $J^{PC}=2^{++}$, the $X(1600)$.
Taking the experimental mass for the threshold $T(M_1,M_2)$ in the light-quark case together with the values obtained for $\Delta$,
one can estimate the energy of these states, being $M(2^{++})\approx 1500$ MeV and $M(1^{-+})\approx 2900$ MeV.
The large mass obtained for the $1^{-+}$ four-quark state makes doubtful 
the identification of the $\pi_1(1400)$ or the $\pi_1(1600)$ with a pure multiquark state, although
a complete calculation is needed before
drawing any definitive conclusion.
Concerning the $X(1600)$, being its experimental mass 1600$\pm$100 MeV, a tetraquark configuration
seems likely.

This work has been partially funded by MCyT
under Contract No. FPA2004-05616, by JCyL under
Contract No. SA104/04, and by GV under Contract No.
GV05/276.

\end{document}